\documentclass{article}
\usepackage{epsfig}
\usepackage{amssymb}
\def\x{{\mbox{\boldmath$x$}}}

\def\r{{\mbox{\boldmath$r$}}}
\def\unitr{{\mbox{\boldmath$\hat r$}}}

\def\v{{\mbox{\boldmath$v$}}}

\newcommand{\be}{\begin{equation}}
\newcommand{\ee}{\end{equation}}
\newcommand{\bea}{\begin{eqnarray}}
\newcommand{\eea}{\end{eqnarray}}

\begin{document}
\title{Isotropy {\it vs} anisotropy in small-scale turbulence}
\author{Luca Biferale$^1$ and Massimo Vergassola$^2$} \maketitle
\centerline{$^1$Dipartimento di Fisica and INFM , Universit\`a "Tor
Vergata",} \centerline{ Via della Ricerca Scientifica 1, I-00133 Roma,
Italy} \centerline{$^2$Observatoire de la C\^ote d'Azur, B.P. 4229,
06304 Nice Cedex 4, France}

\begin{abstract}
The decay of large-scale anisotropies in small-scale turbulent flow is
investigated. By introducing two different kinds of estimators we
discuss the relation between the presence of a hierarchy for the
isotropic and the anisotropic scaling exponents and the persistence of
anisotropies.  Direct measurements from a channel flow numerical
simulation are presented.
\end{abstract}
\vskip 0.5cm One of the main assumptions made by A.N.~Kolmogorov in
his 1941 theory is the restoring of universality and isotropy at small
scales in turbulent flows. The idea is that the effects of a
large-scale anisotropic forcing and/or boundary conditions are rapidly
lost during the process of energy transfer toward the small
scales. The overall result is that the isotropy and the universality
of turbulent fluctuations should be locally restored at small enough
scales and large enough Reynolds numbers. The rate of convergence
toward isotropy can be quantitatively predicted within the K41 theory
both as a function of the scale, e.g.  for the structure functions,
and as a function of the Reynolds numbers, e.g. for the single-point
moments of the velocity gradients.  Experiments \cite{gw98} and
numerical simulations \cite{p96,ps96,se00} do not confirm those
predictions.  The skewness of the transversal gradients, $S_3= \langle
(\partial_yu_x)^3\rangle/\langle(\partial_y u_x)^2\rangle^{3/2}$ is
for example found to have a very slow decay with $Re_{\lambda}$.  The
effect is even stronger for the fifth-order skewness $S_5= \langle
(\partial_yu_x)^5\rangle/\langle(\partial_y u_x)^2\rangle^{5/2}$,
observed to remain $O(1)$ for all available $Re_{\lambda}$.  Similar
results were recently reported on a series of hydrodynamical
problems. The most striking ones were obtained analytically in passive
scalar/vector models advected by isotropic, Gaussian and white-in-time
velocity fields (the so-called Kraichnan model \cite{rhk94}) with a
large scale anisotropic forcing \cite{pas,lm00}. Numerical
\cite{p94,clmv99} and experimental (see, e.g., \cite{Sreeni91,war_pass})
evidences of persistence of anisotropies in real passive scalars have
also been reported.

On one hand, there are then strong indications in favor of a
persistent memory of the large-scale anisotropies even at the smallest
scales of a turbulent flow. On the other hand, there are theoretical
arguments \cite{alp00} going in the opposite direction, i.e.  that
anisotropic fluctuations are sub-dominant with respect to the isotropic
ones (see below). This short note is meant to clarify the relation
between the previous results and support the arguments by numerical
simulations on channel flow turbulence.

The analysis in \cite{alp00} is based on the invariance under
rotations of the unforced Navier-Stokes equations. We shall
specifically restrict here to the structure functions and refer to the
original paper for more complicated tensorial objects.  Since the
Navier-Stokes equations are invariant under rotations, the
correlations are conveniently decomposed in terms of the irreducible
representations of the rotation group.  For the $n$-th order
longitudinal structure function we have for example
\begin{equation}
S_n(\r) = \langle
[((\v(\x)-\v(\x+\r))\cdot \r]^n\rangle=\sum_{jm} S_n^{jm}(|r|) Y_{jm}(\unitr),
\label{so3_sf}
\end{equation} 
where we have explicitly used the fact that the basis of the rotation
group for scalar functions is the set of spherical harmonics
$Y_{jm}$. The coefficients $S_n^{jm}(|r|)$ are expected to behave as
power laws $r^{\xi_n^j}$ and the different scaling exponents to depend
on the index $j$ (the exponents should not depend on $m$ since it does
not appear in the equations of motion --- see \cite{alp00} for more
details). The previous strong assumption is motivated by the idea of
universality, i.e. that inertial-range scaling behaviors are
independent of the large-scale boundary and forcing
effects. Furthermore, it is natural to suppose a hierarchical
organization of the different sectors in the inertial range, i.e. the
existence of a hierarchy among the scaling exponents characterizing
different sectors\,:
\begin{equation}
\xi_n^{j=0} < \xi_n^{j=1} < \xi_n^{j=2} < \cdots
\label{gerarchia2}
\end{equation}
This statement, even if not proved for the Navier-Stokes equations, is
verified analytically in various Kraichnan models of passive fields
\cite{lm99,abp00}.  The existence of the hierarchy (\ref{gerarchia2})
implies that the anisotropic fluctuations become more and more
subdominant at the small scales as their degree of anisotropy
increases.

Let us now analyze in a quantitative way the relative importance of
isotropic and anisotropic fluctuations. In the following we shall
concentrate for simplicity on the structure functions, but the same
arguments could be generalized to other correlations. Isotropic flows
are characterized by having only the sector $j=0,m=0$ excited. One is
therefore naturally lead to introduce two different tests to quantify
the degree of isotropy/anisotropy. First, (case A) one can analyze
fluctuations of comparable intensity, i.e. fixing the order $n$ of the
structure function and measuring the scaling in different sectors. We
can for example introduce the ratio between the projection on the
anisotropic sector with the non-vanishing indices $j,m$ and the
projection on the isotropic sector $j=m=0$\,:
\begin{equation}
 T^{jm}_n(r) \equiv \frac{S_n^{jm}(r)}{S_n^{00}(r)}.
\label{caseA}
\end{equation}
We thus have the possibility to disentangle different degrees of
anisotropy depending on the typical intensity of the velocity
fluctuations. Looking at the structure functions of low order (small
$n$'s) gives a test on the isotropy of the weak fluctuations, while
looking at high orders (large $n$'s) gives a test on the statistics of
strong turbulent fluctuations.  A second possible estimator (case B)
consists in first normalizing the field and then taking moments of
it. As it is done for the skewness, the kurtosis etc etc., we can for
example normalize by the isotropic component of the second order
longitudinal structure function, $S_2^{00}(|r|) =\langle
[((\v(\x)-\v(\x+\r))\cdot \r]^2\rangle_{j=0,m=0}$.  The resulting
dimensionless stochastic variable can then be studied by looking at
its decomposition in different $j,m$ sectors:
\begin{equation}
  \hat{S}_n^{jm}(|r|) \equiv 
\frac{S_n^{jm}(|r|)}{\left(S_2^{00}(|r|)\right)^{n/2}}.
\label{caseB}
\end{equation}

If the hierarchy (\ref{gerarchia2}) holds, all the observables (A)
tend to zero as the scale is decreased. The decay rates possibly
differ from the dimensional predictions due to intermittency, but they
are guaranteed to be positive. There is no experimental or numerical
evidence that the hierarchy (\ref{gerarchia2}) is violated.  The
situation with observable (B) is quite different.  The dimensionless
quantities are indeed formed by comparing anisotropic and isotropic
fluctuations of different intensity (in the numerator and denominator
of (\ref{caseB}) structure functions of different orders are
involved). The hierarchy (\ref{gerarchia2}) does not give any
constraint in this case and it is well possible that $\xi_n^j <
\frac{n}{2} \xi_2^{j=0}$.  The corresponding observable (B)
$\hat{S}_n^{jm}$ defined in (\ref{caseB}) would then diverge going
toward the small scales, even in the presence of the hierarchy
(\ref{gerarchia2}). That divergence is the effect of persistence of
anisotropies reported in experiments and numerical simulations both
for the passive scalars and Navier-Stokes turbulence (see
\cite{pas,war_pass}). It is of importance to notice that the
persistence of anisotropies is a combined effect of anisotropy and
intermittency. If the hierarchy (\ref{gerarchia2}) is valid and there
is not intermittency in all the sectors, i.e.  $\xi_n^j = \frac{n}{2}
\xi_2^j$, then the observables (B) with positive $j$ would vanish at
small scales.

Let us now support the above arguments by presenting some results
obtained in channel flow simulations. The simulations are performed on
a grid of $128\times128\times 256$ points with periodic boundary
conditions in the stream-wise and span-wise directions and no-sleep
boundary conditions at the top and the bottom walls. At the center of
the channel we have $Re_{\lambda} \sim 70$. Due to the relatively
moderate Reynolds number, no scaling laws are observed.  Still, even
in the absence of scaling laws, it is quite clear from the data that
the two sets of observable (A) and (B) behave in a very different way.
In Fig.~1 we present the quantities (A) and (B) for the structure
functions of order 4 and 6 at the center of the channel for the sector
$j=2,m=2$.  In Fig.~2 the same is presented but for a higher sector,
$j=4,m=2$.  While the observable (A) always monotonically decreases
with the scale, the observable (B) of the sixth order shows a clean
tendency to increase. That is the manifestation of the persistence of
anisotropies at the small scales and gives further support to the
observations first made in \cite{ps96}. Note that the scales shown in
the figure go from the largest available one (the box size) to the
beginning of the viscous scale. The decomposition in spherical
harmonics at the very small scales (inside the viscous range) is hard
to obtain numerically because of interpolation errors of the cubic
grid on the sphere.  Details on the numerical procedure to compute the
observable shown here can be found in \cite{abmp}.

As for the intermittency in the anisotropic sectors, the situation is
still moot. There is only one attempt to directly measure the
projections on each single sector in the same channel flow data set
used here \cite{abmp,blmt}. 
As stated previously, the Reynolds number is unfortunately
not high enough and scaling exponents of the anisotropic sectors can
be measured only via the ESS \cite{ess}. In the anisotropic sectors it
is even not quite clear what would be the dimensional prediction for
the $\xi^j_n$ with $j>0$. Different dimensionless quantities can
indeed be built by using some anisotropic mean observable, e.g. the
mean shear, and the usual energy dissipation. The dimensional
predictions would then depend on the requirement that the anisotropic
correction is (or is not) an analytical, smooth deviation from the
isotropic sector. Furthermore, the comparison with the behavior
observed in the Kraichnan models of scalar/vector fields
\cite{rhk94,v96} suggests that the anisotropic sectors may show
intermittent corrections induced by the homogeneous (non-linear, in
the Navier-Stokes case) part of the equations for the correlation
functions.  If that is the case, the dimensional predictions might be
very far from the observed behaviors.

In conclusion, we have discussed the decay of large-scale anisotropy
memory in the small scales of turbulent flows. The analysis of
numerical data from channel flow simulations indicate that the
anisotropies persist at the small scales but still respecting the
hierarchy (\ref{gerarchia2}) between the isotropic and anisotropic
velocity components.

We are grateful to A.~Mazzino, I.~Procaccia, A.~Pumir and B.~Shraiman
for many stimulating discussions.  This research was supported in part
by the European Union under contract HPRN-CT-2000-00162 and by the
National Science Foundation under Grant No. PHY94-07194.

\newpage
\centerline{FIGURE CAPTIONS}
\noindent 
{\bf FIGURE  1}: Analysis of the persistence of anisotropies with observable
belonging to case A and B for the projection of structure functions in
the sector $j=2,m=2$. Bottom: projections of fourth moment,
$T_4^{2,2}(|r|)$ ($\times$) and $\hat{S}_4^{2,2}(|r|)$ ($+$). Top:
projection of sixth moment, $T_6^{2,2}(|r|)$ (Squares) and
$\hat{S}_6^{2,2}(|r|)$ ($\ast$). Notice how for the moment of order
$6$ we have a clear tendency toward increasing of
$\hat{S}_6^{2,2}(|r|)$ at small scales. Scales are dropped at $R \sim
10$ which corresponds to the onset of the viscous scale in the
simulation.  For details on how to compute numerically the
projections, $S_n^{jm}(|r|)$ see \cite{abmp}\\
\vskip 1 truecm
\noindent 
{\bf FIGURE 2}: The same of figure 1 but for the sector
$j=4,m=2$. Bottom: projections of fourth moment, $T_4^{4,2}(|r|)$
($\times$) and $\hat{S}_4^{4,2}(|r|)$ ($+$). Top: projection of sixth
moment, $T_6^{4,2}(|r|)$ (Squares) and $\hat{S}_6^{4,2}(|r|)$ ($\ast$).
\newpage
\centerline{FIGURE 1, L. Biferale and M. Vergassola}
\begin{figure}
\epsfxsize=8truecm
\centerline{\rotatebox{270}{\epsfbox{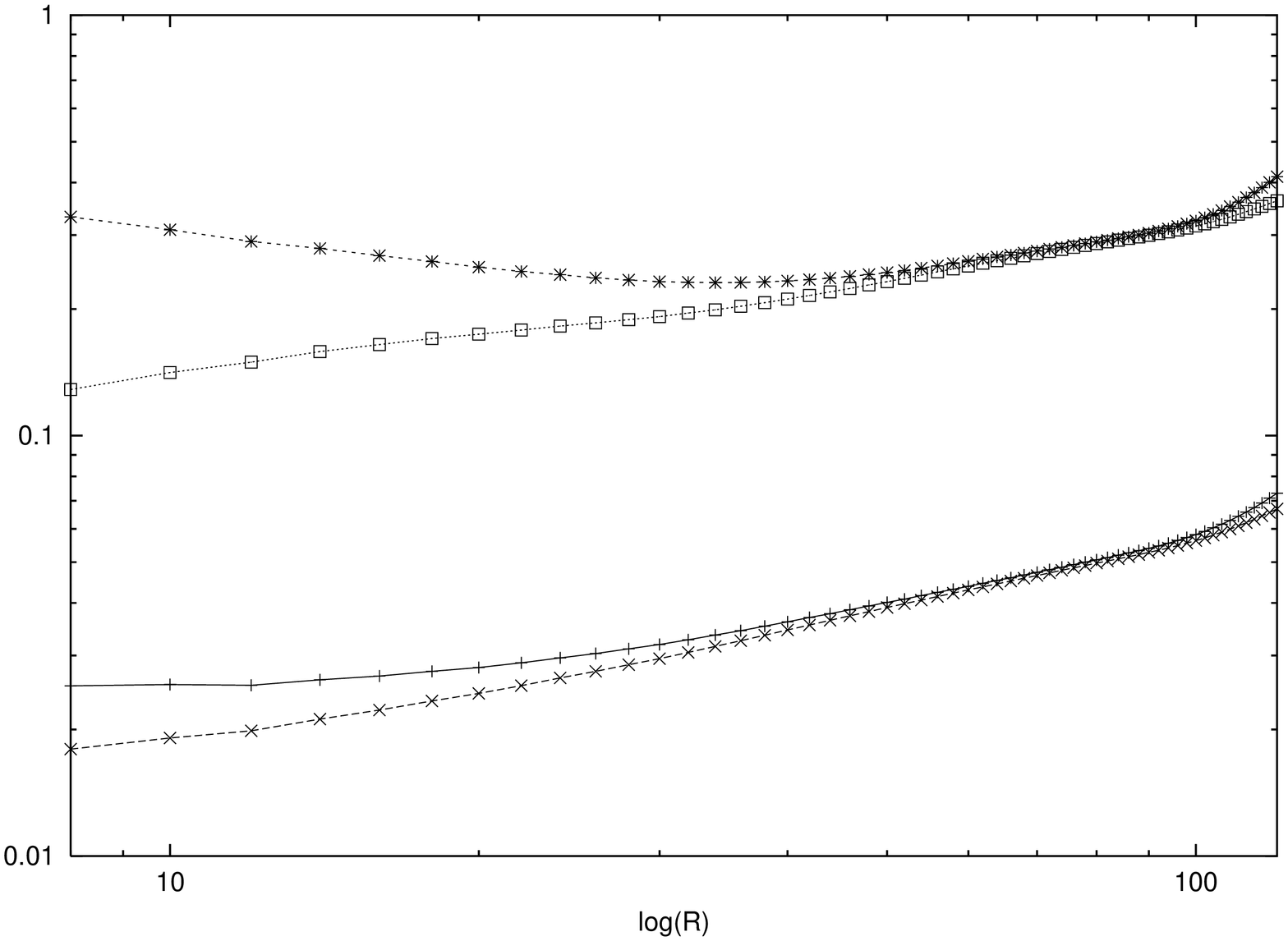}}}
\label{fig1}
\end{figure}
\newpage 
\centerline{FIGURE 2, L. Biferale and M. Vergassola}
\begin{figure}
\epsfxsize=8truecm
\centerline{\rotatebox{270}{\epsfbox{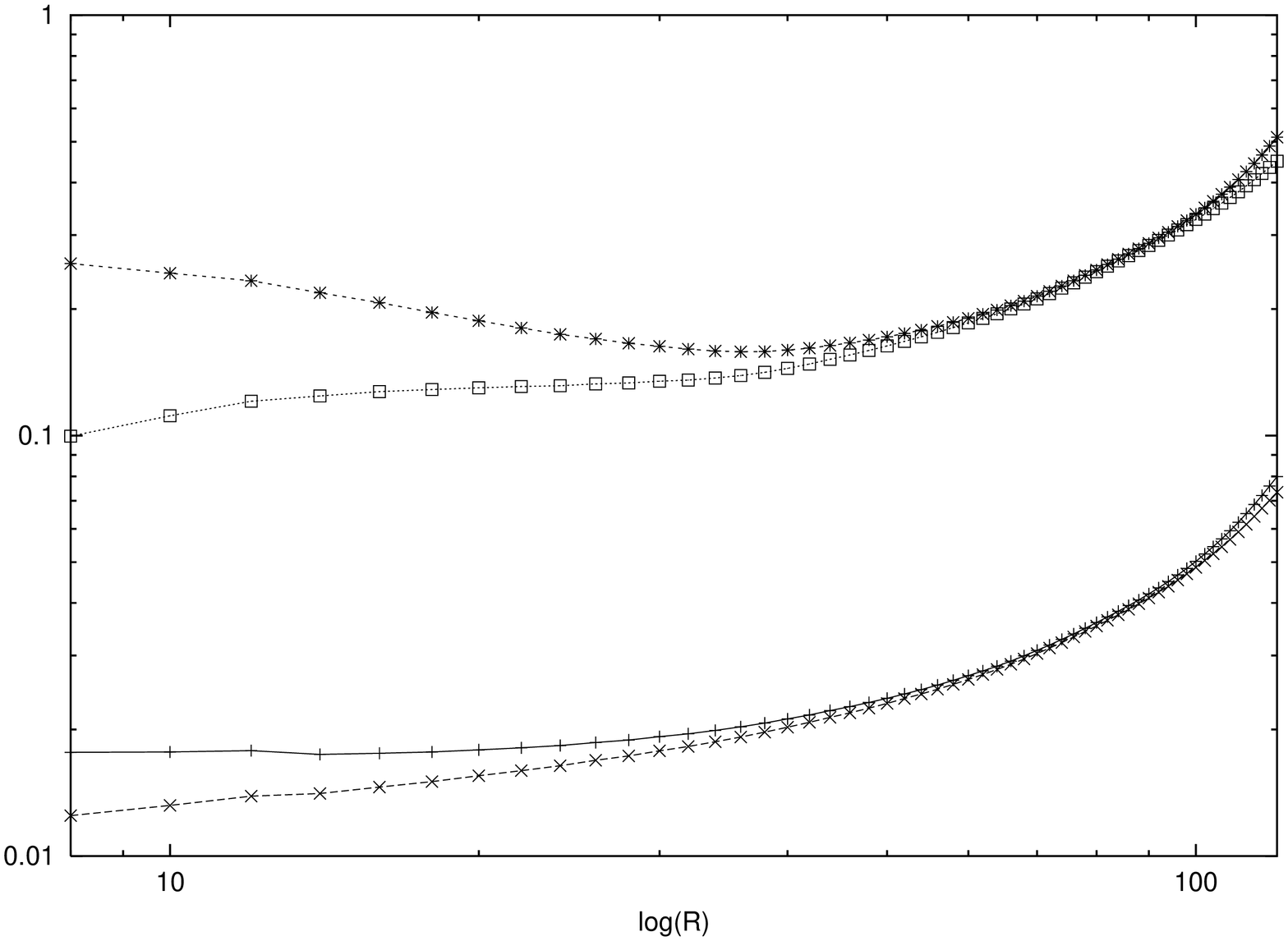}}}
\label{fig2}
\end{figure}


\begin{thebibliography}{99}

\bibitem{gw98} S. Garg and Z. Warhaft, ``On the small scale structure
of simple shear flow'', Phys. Fluids, {\bf 10} 662 (1998).

\bibitem{p96} A. Pumir, ``Turbulence in homogeneos shear flow'' 
 Phys. Fluids {\bf 8} 3112 (1996).

\bibitem{ps96} A. Pumir and B. Shraiman, ``Persistent small scale
anisotropy in homogeneous shear flow''  Phys. Rev. Lett. {\bf 75}
3114 (1996).

\bibitem{se00} J.~Schumacher and B.~Eckhardt, ``On statistically
stationary homogeneous shear turbulence'', nlin.CD/0005021, (2000).

\bibitem{rhk94} R.H.~Kraichnan, ``Anomalous scaling of a randomly
advected passive scalar,'' Phys. Rev. Lett. {\bf 72}, 1016 (1994).

\bibitem{pas} B.I.~Shraiman and E.D.~Siggia, ``Scalar turbulence,''
Nature {\bf 405}, 639 (2000).

\bibitem{lm00}N. V. Antonov, A. Lanotte and A. Mazzino ``Persistence
of small-scale anisotropies and anomalous scaling in a model of
magnetohydrodynamics turbulence'' Phys. Rev. E {\bf 61} 6586 (2000).

\bibitem{p94} A.~Pumir, ``A numerical study of the mixing of a passive
scalar in three dimensions in the presence of a mean gradient,''
Phys. Fluids {\bf 6}, 2118 (1994).

\bibitem{clmv99} A.~Celani, A.~Lanotte, A.~ Mazzino and
M.~Vergassola, ``Universality and saturation of intermittency in
passive scalar turbulence,'' Phys. Rev. Lett. {\bf 84}, 2385 (2000)
 
\bibitem{Sreeni91} K.R.~Sreenivasan, ``On local isotropy of passive
scalars in turbulent shear flows,'' Proc. Roy. Soc. London {\bf A434},
165, (1991).
 
\bibitem{war_pass} Z. Warhaft, ``Passive Scalars in Turbulent Flows''
Annu.  Rev.  Fluid Mech. {\bf 32} 203 (2000).

\bibitem{alp00} I. Arad,  V. L'vov and I. Procaccia, ``Correlation functions
in isotropic and anisotropic turbulence: The role of the symmetry group'',
Phys. Rev. E {\bf 81} 6753 (1999).

\bibitem{lm99} A. Lanotte and A. Mazzino, ``Anisotropic nonperturbative
zeromodes for passively advected magnetic fields'' Phys. Rev. E 
{\bf 60} R3483 (1999).

\bibitem{abp00} I. Arad, L. Biferale and I. Procaccia\,;
``Nonperturbative spectrum of anomalous scaling exponents in the anisotropic 
sectors of passively advected magnetic fields''
Phys. Rev. E {\bf 61}, 2654, 2000.

\bibitem{abmp}  I. Arad, L. Biferale, I. Mazzitelli and I. Procaccia,
``Disentangling scaling properties in anisotropic and inhomogeneous
Turbulence'' Phys. Rev. Lett. {\bf 82} 5040 (1999).


\bibitem{blmt}   L. Biferale, D. Lohse, I. M. Mazzitelli, F. Toschi,
``Probing structures in channel flow through SO(3) and SO(2) decomposition''
 nlin.CD/0006031, (2000).

\bibitem{ess} R. Benzi, S. Ciliberto, R. Tripiccione, C. Baudet, F. Massaioli
and S. Succi, ``Extendend Self Similarity in Turbulent
Flows'' Phys. Rev. E {\bf 48} R29 (1993).

\bibitem{v96} M. Vergassola, ``Anomalous scaling for passively
advected magnetic fields'' Phs. Rev. E {\bf 53} R3021 (1996)

\end{thebibliography}
\end{document}